\documentstyle[twoside,12pt]{article}
\newcommand{\nc}{\newcommand}
\nc{\la}{\lambda} \nc{\La}{\Lambda}  \nc{\al}{\alpha}
\nc{\te}{\theta}  \nc{\be}{\beta}
\nc{\ga}{\gamma}  \nc{\Ga}{\Gamma}
\nc{\de}{\delta}  \nc{\De}{\Delta} 
\nc{\si}{\sigma}  \nc{\ka}{\kappa}
\nc{\om}{\omega}  \nc{\Om}{\Omega}
\nc{\nf}{\infty}  \nc{\qq}{\quad\quad}
\nc{\nl}{\newline\noindent}
\nc{\dst}{\displaystyle}
\nc{\ra}{\longrightarrow}  
\nc{\beq}{\begin{equation}}  \nc{\eeq}{\end{equation}}
\nc{\beqa}{\begin{eqnarray}} \nc{\eeqa}{\end{eqnarray}}  
\nc{\nnb}{\nonumber}
\nc{\brm}{\begin{remunerate}} \nc{\erm}{\end{remunerate}}     
\title {\bf Non linear $\si$ models : \\  renormalisability versus
geometry.\thanks {\noindent Talk given in Cortona, Italy, at the
$1^{st}$ June meeting in honnor of Carlo Becchi $60^{th}$ birthday.}}
\begin{document}
\author{Guy Bonneau\thanks {\noindent Laboratoire de Physique Th\'eorique
et des Hautes Energies,
 Unit\'e associ\'ee au CNRS UMR 7589,``Universit\'e Paris 7,
 2 Place Jussieu, 75251 Paris Cedex 05.\nl\noindent 
bonneau@lpthe.jussieu.fr }}
\maketitle
\begin{abstract}
\noindent After some recalls on the standard (non)-linear $\si$
model, we discuss the interest of B.R.S. symmetry in non-linear $\si$
models renormalisation. We also emphasise the importance of a correct
definition of a theory through physical constraints rather than as given
by a particular Lagrangian and discuss some ways to enlarge the notion of
renormalisability. 
\end{abstract} 

\vfill {\bf PAR/LPTHE/99-27/hep-th/xxx}\hfill  July 99
\newpage

\section{Introduction}

The subject of non-linear $\si$ models is relevant to this meeting for two
reasons :
\begin{itemize}\item Carlo Becchi has been involved in that field in
several occasions. A tentative list could be :
\begin{itemize}\item Axial anomaly in a linear $\si$ model with fermions
(1973) \cite{CB73},
\item Non-linear $\si$ model and the bosonic string (1986-1988)
\cite{CB86},
\item Renormalisability and I.R. finiteness in non-linear $\si$ models on
coset spaces (1987-1988) \cite{CB88},
\item Non-linear $\si$ model with Wess-Zumino term (1989) \cite{CB89},
\item Non-linear $\si$ model with (4,0) supersymmetry (1990) \cite{CB90}.
\end{itemize}
\item In Paris, with Galliano Valent and Fran\c{c}ois Delduc \cite{BDV},
between 1983 and 1987 we studied the quantisation of non-linear $\si$
models in 2 space-time dimensions for a large class of target spaces
including
$CP^1,\ CP^n,\
$the grassmannians, K\"ahler symmetric then K\"ahler homogeneous ones.
After his sabbatical year in Paris in 1981, we kept contact with Carlo
and his group and I had the chance to suggest a collaboration between our
two groups, which led to the renormalisation of all coset-space models
with the algebraic methods {\it \`a la} B.R.S. \cite {CB88}.   My
personal interest in that field was mostly its use as a toy model to
discuss new theoretical ideas such as on shell renormalisation and the
enlargment of the notion of what a renormalisable theory might be. 
\end{itemize}

\noindent In the first part of this talk, I will recall some well known
facts on  non-linear $\si$ models whose physical interest goes back to
current algebra, P.C.A.C., broken symmetries and Goldstone mechanism.
Other fields such as statistical models in $2 + \epsilon$
dimensions, 1/N non perturbative methods, mass gap computations or
string vacuum description, will not be considered (see however the talk
of Fran\c{c}ois Delduc
\cite{FD99}).

In a second part, I will explain the power of B.R.S. methods in that
field, the possible ``apparent" non-renormalisability and end with some
open questions on ``generalised renormalisability".

\section{The ``old" O(N+1) $\si$ model.}
\subsection{The linear $\si$ model.} In their famous paper, GellMann and
L\'evy \cite{GML60} present several Lagrangians ensuring a realisation of
the Partially Conserved  Axial Current (P.C.A.C.) observation ; one of
them was proposed by Schwinger \cite{S57} and writes :
\beq\label{a1} {\cal L} = (1/2)[(\partial_{\mu}\vec{\pi})^2 +
(\partial_{\mu}\si)^2] - (1/2)m^2[\vec{\pi}^2 + \si^2] +
(\la/4)[\vec{\pi}^2 + \si^2]^2 + {\cal L}_{\,breaking}\ ,
\eeq where the breaking of $O(N+1)$ to $O(N)$ is given by a non vanishing
$\si$ field expectation value, resulting from a linear 
\beq\label{a2} {\cal L}_{\,breaking} = c\ \si\,.\eeq The generators of
the symmetry being respectively :
\beq\label{a3} W_{ij}^V  \sim  \int dx\left[\pi_i\frac{\de}{\de\pi_j} -
\pi_j\frac{\de}{\de\pi_i}\right]\ ,\ \ W_{k}^A  \sim \int
dx\left[\si\frac{\de}{\de\pi_k} -
\pi_k\frac{\de}{\de\si}\right]\,,\eeq the $O(N+1)$ algebra is :
\beqa\label{a4}
\left[W^V_{ij},\,W^V_{kl}\right] & = & \de_{jk}W^V_{il} -
\de_{jl}W^V_{ik} -
\de_{ik}W^V_{jl} + \de_{il}W^V_{jk}\,, \nnb\\
\left[W^V_{ij},\,W^A_{k}\right] & = &
\de_{jk}W^A_{i} - \de_{ik}W^A_{j}\,,\\ \left[W^A_{i},\,W^A_{j}\right] & =
& - W^V_{ij}\,.\nnb
\eeqa In presence of the term (\ref{a2}), the ``axial" current
conservation is broken according to the strong P.C.A.C.- Adler condition
:\beq\label{a5} J^{A\,i}_{\mu}(x) = \si\partial_{\mu}\pi^i -\pi^i
\partial_{\mu}\si\ \,,\ \ \partial_{\mu}^x<J_{\mu}^{A\,i}(x) X> =
-c<\pi^i(x) X> +\ {\rm contact\ terms}\,.\eeq Note that, in a nice paper
of 1973 \cite{CB73}, Carlo Becchi gave the first complete analysis of a
linear $\si$ model with fermions, defined by the requirement of the
P.C.A.C.-Adler condition (\ref{a5}).

\subsection{The non-linear $S^N$ model.} In their paper, GellMann and
L\'evy also proposed another Lagrangian, taking into account the
experimental absence of any particle
$\si$  similar to the pions ; moreover, it also explains the smallness
of the mass of the pions when compared to others hadrons. Thanks to the
constraint $$\vec{\pi}^2 + \si^2 = f_{\pi}^2\,,$$
the
$O(N+1)$ symmetry  is now non-linearly realised
\cite{NLS}. The algebra (\ref{a4}) is left unchanged but the 
$W_i^A$ generators are now non-linear ones thanks to the replacement $\si
= \sqrt{f_{\pi}^2 - \vec{\pi}^2}\,.$ The pion field
belongs to the symmetric space $S^N = O(N+1)/O(N)\,.$ The Lagrangian
(\ref{a1}) now describes massless particles. The non zero mass of the
physical pions - and the correlative non exact conservation of the axial
current - will result either from an explicit breaking $$\De{\cal L} =
c\sqrt{f_{\pi}^2 - \vec{\pi}^2}\,,$$ or from a Nambu-Golstone mechanism,
through the spontaneous breaking of the
$O(N+1)$ symmetry to the ``vector"
$O(N)$ one. The ``axial" symmetry is now non-linearly realised, according
to :
\beq\label{b1}
\de_{\,\be}\vec{\pi} = \vec{\be}\sqrt{f_{\pi}^2 - \vec{\pi}^2}\,,\ \
\de_{\,\ga}\de_{\,\be}\vec{\pi} = -\vec{\be}(\vec{\ga}.\vec{\pi}) \,,\eeq
and the axial current is :
\beq\label{b2} J^{A\,i}_{\mu}(x) = \sqrt{f_{\pi}^2 - \vec{\pi}^2}
\partial_{\mu}\pi^i -\pi^i
\partial_{\mu}\sqrt{f_{\pi}^2 - \vec{\pi}^2}\ \ {\rm with}\ \
\partial_{\mu}J^{A\,i}_{\mu}(x) = -c\,\pi^i(x)\,.
\eeq At the end of the sixties, physicists use the geometry of coset
spaces
$G/H$ to study various realisations of the chiral symmetry [ understood
as being  rather $\frac{SU(2)\times SU(2)}{SU(2)}$ than
$\frac{O(4)}{O(3)}$ ] ; later on, this geometrical approach was
generalised, in the framework of 2-dimensional field theory, to an
arbitrary riemmannian manifold (see 
\cite{F} for a review), as I  recall in the next subsection.

\subsection{The non-linear $\si$ model.} Given an arbitrary N-dimensional
 riemannian space, the lagrangian density is \footnote{\, Here
we do not consider a Wess-Zumino term in the Lagrangian ( in two
space-time dimensions, antisymmetric tensors $\epsilon^{\mu\nu}$ and
$b_{ij}$ would respectively replace $\eta^{\mu\nu}$ and $g_{ij}\,).$ Note
that Carlo Becchi has also been involved in that situation ( in a work
with Olivier Piguet
\cite{CB89}).} written :
\beq\label{c1} {\cal L} = (ds)^2 = \frac{1}{2}\left[
\eta^{\mu\nu}g_{\,ij}(\phi)\partial_{\mu}\phi^i
\partial_{\nu}\phi^j\right]\,,\eeq where \begin{itemize}
\item the fields $\phi^i$ are considered as coordinates on the manifold
(the target space),
\item $ g_{\,ij}(\phi)$ is a metric on the manifold, 
\item $\eta^{\mu\nu}$ is the space-time (inverse)metric.\end{itemize}
When isometries are present - for instance when the manifold is an
homogeneous one (coset space G/H) -, they may be realised in various ways
:
\beq\label{c2}
\left[W_i^{G/H},\,\phi^j\,\right] = - i f_{ij}(\phi)\, ;\eeq 
$f_{ij}(\phi)$ is constrained by the Lie algebra structure and, thanks to
the invariance of the distance, related to the target space metric. For
example, in the $S^N$ case, the standard parametrisation gives :
$$ f_{ij}(\vec{\pi}) = \de_{ij}\sqrt{f_{\pi}^2 - \vec{\pi}^2}\,,\ \
g_{ij}(\vec{\pi}) = \de_{ij} + \frac{\pi_i\pi_j}{f_{\pi}^2 -
\vec{\pi}^2}\ \,;$$ the stereographic parametrisation of Schwinger
\cite{S1} gives :
$$f_{ij}(\vec{\phi}) = \frac{1}{2}\de_{ij}(f_{\pi}^2 - \vec{\phi}^2) +
\phi^i\phi^j\,,\ \ g_{ij}(\vec{\phi}) = \frac{\de_{ij}}{ (f_{\pi}^2 +
\vec{\phi}^2)^2}\ \,; $$ note that the change of coordinates between these
two  parametrisations is non-linear :
$$\vec{\phi} = \frac{f_{\pi} \vec{\pi}}{f_{\pi}  + \sqrt{f_{\pi}^2 -
\vec{\pi}^2}} \Leftrightarrow \vec{\pi} = \frac{2
f_{\pi}^2\vec{\phi}}{(f_{\pi}^2 + \vec{\phi}^2)}\,.$$ Other choices have
been considered, for example one where $\det\mid\mid g_{ij}\mid\mid = 1$
( Charap
\cite{C1}), with the hope that they will lead, to all-order of
perturbation theory, to finite S-matrix elements in 4 dimensions.
\newline 

Let us end that Section by some comments.
\subsection{Remarks and comments.}
\begin{itemize}

\item As physics should be independent of the
parametrisation, in those days some people consider the choice of
parametrisation as a kind of gauge choice \cite{C1} : this idea was
later on taken up in the framework of the use of B.R.S. symmetry in 
non-linear
$\si$ models and non-linear field redefinitions \cite{Ring87}.
\nl Notice that in a recent work, Blasi et al.
\cite{BLASI99} prove the equivalence theorem - independence of S matrix
elements under reparametrisations  of the fields - by that very idea :
the new terms in the Lagrangian density, corresponding to a change of
gauge,  are B.R.S. variation of something, and then  unphysical ones.

\item A perturbative approach requires the choice of a special point ``S"
on the manifold ${\cal M}$ ({\it e.g.} the south pole for stereographic
parametrisation). However, if the manifold is an homogeneous one, all
points are equivalent. On the other hand, in the case of generalised
non-linear
$\si$ models defined on an arbitrary target space, this independence
results from the existence at the quantum level of some compatibility
conditions ensuring the existence of a globally defined metric on ${\cal
M}$, as pointed out by Friedan
\cite{F} and thoroughfully discussed by Carlo Becchi at the 1987
Ringsberg meeting
\cite{CB87}.
\item Given some point ``S" on ${\cal M}$, it would be simpler to have a
covariant perturbative expansion : as first shown by Meetz, normal
coordinates give such a geometric formulation \cite{Meetz}. At the
one-loop order, the divergences are proportional to the Ricci tensor of
the target space metric :
$$\De{\cal L} \sim \de^4(0)R_{icc_{ij}}[g(\phi)]\partial_{\mu}\phi^i
\partial_{\mu}\phi^j\,.$$ For any symmetric space the  Ricci tensor is
proportional to the metric tensor, then the corresponding models are
one-loop renormalisable. Other situations, as well as an all-order
analysis, will be discussed in the next Section. 
\item I would like to emphasize that, in the sixties, physicists tried to
take the P.C.A.C. - Adler condition (\ref{a5}) as {\bf the very
definition of a right theory}. They considered the Lagrangians (\ref{c1}) 
as {\bf effective lagrangians} and agreed on the importance of the algebra
of isometries - the Ward identities. For example, many efforts \cite{C2}
were done to enforce P.C.A.C. in higher orders of perturbation theory (in
4 space-time dimensions). 
\end{itemize}
\noindent It seems to me that after 1972 or so, there appeared some confusion :
I feel that the successes of dimensional regularisation for gauge
theories - with the kind of ``magic" that accompanies its use - and the
discovery of the ``standard model" with its simple Lagrangian, somehow
obscure the comprehension ; so, in the rest of this talk, I shall comment
on the definition and renormalisation of non-linear
$\si$ models, this time in two space-time dimensions where they are power
counting renormalisable, the canonical dimension of the fields being zero.

\section{From Ward identities to B.R.S. symmetry.} In the presence of
non-linear transformations of the fields, the standard trick is to add
to the effective action sources for these variations, as well as for
their  successive iterations. In doing so, one may need an infinite series
of sources. This makes the quantum analysis {\it a priori} a tremendous
task : however, in some cases   it  may be carried out :
\begin{itemize}\item when there exists a special parametrisation of the
coset space such that the series of transformations closes in a finite
number of steps. This is the case for $S^N$ in the standard
parametrisation, as well as for the Grassmannians \cite{Valent1} ;
\item for K\"ahler coset spaces, there is a $U(1)$ charge, which helps to
control the infinite tower of transformations. In particular, our group in
Paris was able to exhibit a parametrisation where the isometries are
holomorphic :
$$\de\phi^i = \epsilon^i + F^i_{jk}\epsilon^j\phi^k +
G^i_{\bar{j}k}\bar{\epsilon}^{\bar{j}}\phi^k\ \ ,\ \ 
\de\bar{\phi}^{\bar{i}} = \bar{\epsilon}^{\bar{i}} +  F^{\bar{i}}_{\bar{j}
\bar{k}}\bar{\epsilon}^{\bar{j}}\bar{\phi}^{\bar{k}} +
G^{\bar{i}}_{j\bar{k}}\epsilon^{j}\bar{\phi}^{\bar{k}}\,,$$ and so to give
a complete all order analysis of the non-linear $\si$ models on
homogeneous K\"ahler spaces \cite{BDV2}.\end{itemize} 
\noindent But, if one recalls that the algebra involves commutators
rather than products of transformations and that for coset spaces the
commutator of two infinitesimal transformations gives another
infinitesimal group transformation, it is tempting to promote the
parameters of the transformations to anti-commuting ones (Fadeev-Popov
constant ghosts). This idea was proposed by Alberto Blasi and Renzo
Collina for the $S^N$ model \cite{BC87}, and we used it in our
Genoa-Paris collaboration to obtain a complete algebraic proof of the
renormalisability of any homogeneous non-linear $\si$ model
\cite{CB88}. Let us remark that :\begin{itemize}\item  we replace
Ward Identities by {\bf B.R.S. invariance and Slavnov identities}. No
parametrisation being now specified, it is not surprising that the field
may be non-linearly renormalised \footnote{\, Contrarily to the
homogeneous K\"ahler case where in our parametrisation, using Ward
identities, we prove that the field is not renormalised ($Z_{\phi} =
1$)}. The B.R.S. invariant effective action writes :
\beq\label{g1}
\Ga^{(0)} = S \De^{(-1)} + A^{(0)}\,,\eeq where $S$ is the Slavnov
operator,
$A^{(0)}$ the invariant action and $\De^{(-1)}$ an arbitrary  local
functional in the fields and their derivatives, constrained by power
counting arguments and of Fadeev-Popov charge -1 ;
\item this {\bf first extension of the notion of renormalisability} (as an
infinite number of renormalisation constants appear) was first exhibited
by Piguet and Sibold in supersymmetric models \cite{PS82} and widely
discussed at the 1987 workshop in Ringsberg \cite{Ring87}.\end{itemize}

\noindent During this workshop,  Breitenlohner and Maison exposed their efforts
to quantise N=2 super Yang-Mills theory \cite{BM87} : on the one hand,
they came up against the difficulty of an infinite series of sources and
realized that this B.R.S. idea will reveal itself very fruitful. On the
other hand, they also came up against the difficulty of open algebras
that occur in supersymmetric theories (the algebra only {\bf  closes
on-shell} and, moreover, in their situation, {\sl modulo} a gauge
transformation). This problem of on-shell closed algebras was tackled by
Kallosh
\cite{Kallosh}, solved through the formalism of Balatin and Vilkovisky
\cite{BV} and exemplified by Piguet and Sibold in their analysis of
Wess-Zumino model without auxiliary fields \cite{PS85}.

This offers a {\bf new success of B.R.S. formalism} as one may define a
modified Slavnov operator and a modified effective action that take care
of that non-closure of the algebra. The challenge of Breitenlohner and
Maison was finally answered by Maggiore \cite{Nicola} ; moreover, this
method was also used in the study of N = 2 and N = 4 supersymmetric
non-linear $\si$ models \cite{B94} and recently applied by Blasi and
Maggiore to broken algebras \cite{BM99}.

\section{On some apparent non-renormalisabilities.} In that Section, I
want to illustrate on concrete examples, two problems that result from
too naive a use of dimensional regularisation.

\subsection{Non-linear renormalisation of the fields.} Take the $O(N)$
invariant Lagrangian : \beq\label{f1} {\cal L}_c =
\frac{1}{2g^2}\left[\frac{(\partial_{\mu}\vec{\phi})^2}{(1 + \vec{\phi}^2
)^2} - m^2 \frac{\vec{\phi}^2}{(1 + \vec{\phi}^2 )}\right]\,.
\eeq Dimensional regularisation gives as 1-loop effective lagrangian :
\beq\label{f2}{\cal L}_c + \De{\cal L}_{min.} = \left[1 + \frac{\hbar
g^2}{\pi\epsilon}\right]{\cal L} + \frac{(N-2)}{4g^2}\frac{\hbar
g^2}{\pi\epsilon}\left[\frac{(\partial_{\mu}\vec{\phi})^2}{(1 +
\vec{\phi}^2 )} - 2 \frac{(\vec{\phi}.\partial_{\mu}\vec{\phi})^2}{(1 +
\vec{\phi}^2 )^2}\right]\,.
\eeq Using standard arguments, {\bf one would conclude that for 
$\boldmath N \neq 2$, the model is non-renormalisable}. This would be
rather unpleasant as the Lagrangian (\ref{f1}) is the ordinary $S^N$ one
in the stereographic parametrisation (subsect. 2.3). As shown in
\cite{B1}, this is a special case of on-shell
renormalisability, as the S-matrix elements can be  renormalised through a
coupling and mass renormalisation :
$$ Z_{g^2} = 1 - \frac{(N-1)\hbar g^2}{\pi\epsilon}\ ,\ \ Z_{m^2} = 1 -
\frac{(N-2)\hbar g^2}{2\pi\epsilon}\,.$$ Indeed, result (\ref{f2}) may be
understood as a non-linear  renormalisation of the fields according to 
:\beq\label{f3}
\vec{\phi}_0 = \vec{\phi} \left[ 1 - \frac{(N-2)\hbar g^2}{4\pi\epsilon} 
(1+\vec{\phi}^2)\right]\,.\eeq 

\subsection{Unsufficient definition of the model.}
 Consider another $O(N)$ invariant Lagrangian : \beq\label{h1} {\cal L}_c
=
\frac{1}{2g^2}\left[\frac{(\partial_{\mu}\vec{\phi})^2}{(1 - \vec{\phi}^2
)} - m^2 \vec{\phi}^2 \right]\,.
\eeq  A calculation with dimensional regularisation give as 1-loop
effective lagrangian :
\beqa\label{h2}{\cal L}_c + \De{\cal L}_{min.} &  = &
\frac{1}{2g^2}\frac{(\partial_{\mu}\vec{\phi})^2}{(1 - \vec{\phi}^2
)}\left[1 -  \frac{\hbar g^2}{2\pi\epsilon}\frac{1}{(1 - \vec{\phi}^2
)}\right]  - \frac{m^2}{2g^2} \vec{\phi}^2 \left[1 + \frac{\hbar
g^2}{2\pi\epsilon}\right] -\nnb \\ & - & \frac{(N-2)}{4g^2}\frac{\hbar
g^2}{2\pi\epsilon}\left[\frac{(\partial_{\mu}\vec{\phi})^2}{(1 -
\vec{\phi}^2 )} + m^2\vec{\phi}^2 +
\frac{(\vec{\phi}.\partial_{\mu}\vec{\phi})^2}{(1 -
\vec{\phi}^2 )^2}\right]\,.
\eeqa
\begin{itemize}\item For N = 2, the model appears to be 1-loop
renormalisable with :$$Z_{\phi} = 1 - \frac{\hbar g^2}{2\pi\epsilon}\,,\
Z_{m^2} = 1 +\frac{\hbar g^2}{\pi\epsilon}\ {\rm and}\ Z_{g^2} = 1\,,$$
which means a vanishing
$\beta$ function at that order \cite{Maillet}. This is fine, but, at
2-loop order, still in the minimal dimensional scheme, the calculation
gives non-renormalisability. It is then tempting to try an ${\cal
O}(\hbar)$ additive modification of the Lagrangian : as shown in
\cite{BD2}, renormalisability is recovered, {\bf but the candidate 1-loop
Lagrangian is not completely determined} : this signals an unsufficient
definition of the model !
\begin{itemize}\item {\bf The solution :} for N = 2, the  classical action
satisfies the factorisation properties of an integrable model (Complex
Sine-Gordon
\cite{PLR}). The {\bf model should be defined by the non-production
physical property}, which means an infinite number of identities.
With Fran\c{c}ois Delduc, we prove that, enforcing one-loop
factorisability, the
${\cal O}(\hbar)$ finite counterterms are completely fixed and as an extra
bonus we get the vanishing of the $\beta$ function at 2-loops order
\cite{BD2}. \item {\bf Comments :}\begin{itemize}

\item As the factorisation property is specific of 2-dimensional
space-time, it is not surprising that dimensional regularisation leads to
quantum corrections, often called ``spurious anomalies".
\item In the eighties, many uncorrect claims were published - for example
on a possible  violation of the Adler-Bardeen theorem in
Super-Yang-Mills. As shown in \cite{BB90}, most of them were due to the
identification of (minimal) regularisation with a renormalisation scheme.
The ``magic" of minimal dimensional regularisation leads to a blind faith
understanding of the Lagrangian as {\bf the theory}, as opposed to
people involved in soft pions and chirality business in the sixties. 
\end{itemize}
\end{itemize}
\noindent This is the right moment to recall a nice expression of Carlo
Becchi :
\newline

\hspace{8cm} {\sl ``The Lagrangian is an opinion"...}

\item For $N > 2$, the model is 1-loop non-renormalisable, even on-shell,
as S-matrix elements cannot be made finite through some renormalisation of
the parameters of the classical Lagrangian (\ref{h1}). This should be
related to the absence, in ordinary perturbation theory, of O(N)
symmetric integrable models.
\end{itemize}

\section{Concluding remarks : generalised renormalisability ?}

In the previous subsections, we discussed how the study of non-linear
$\si$ models, taken as ``toy" models, leads to ``trivial " extensions of
the notion of renormalisability : possible non linear renormalisations of
the fields and on-shell renormalisability. In these concluding Section, I
comment on two other directions :
\begin{itemize}
\item {\bf Friedan's approach \cite{F}} :\nl Friedan discusses
renormalisation ``in the space of metrics", which, in the absence of
isometries, {\sl a priori} leads to some unpredictiveness as a theory
with an arbitrary metric
$g_{ij}[\phi]$ on the manifold ${\cal M}$ would involve an infinite
number of parameters. If only a finite number of them are ``physical"
ones, the theory is acceptable and may be called a renormalisable theory.
Of course, the important part lies in the definition of what are
``physical" parameters. For instance, does a geometrical constraint
(Ricci flatness, hyperk\"ahlerness,..) sufficiently reduce the space of
parameters for a given manifold ? A complete answer to that question is
not known yet.
\item {\bf The effective action point of view \cite{GW96}} : \nl the
B.R.S. methods have been used in a series of papers \cite{HKY} where it
is claimed that the goal of the physicists of the sixties \cite{C2} -
that is to say the validity, to all orders of perturbation theory, of the
low energy theorems for non-linear $\si$ models, in D = 4 dimensions -
has been reached. In another spirit, Gomis and Weinberg  \cite{GW96} note
that, given a standard renormalisable theory, the integration over some
massive fields leads to an Action which is power counting
non-renormalisable. However, they argue that such a theory is physically
predictive and then  should be considered as ``renormalisable" in an
extended sense. Moreover  they discuss the {\sl a priori} constraints
that should be imposed on the form of the bare action to make possible
the absorbtion of any infinity coming from loop graphs by an allowed
counterterm. Here also, more work is to be done.
\end{itemize}
$$   $$
{\bf Aknowledgments :} I sincerely thank Alberto Blasi and Renzo
Collina, the organisers of the meeting for their invitation
 : they offered the pleasure of spending
some time all together in the very nice town of Cortona.
\bibliographystyle{plain}

\begin{thebibliography}{49}

\bibitem{CB73} C. Becchi, {\sl Comm. Math. Phys.} {\bf 33} (1973) 97.

\bibitem{CB86} L. Baulieu, C. Becchi and R. Stora,  {\sl Phys. Lett.}
{\bf 180B} (1986) 55 ; C. Becchi, {\sl Nucl. Phys.} {\bf B304} (1988) 513 

\bibitem{CB88} C. Becchi, A. Blasi, G. Bonneau, R. Collina and F. Delduc,
{\sl Comm. Math. Phys.} {\bf 120} (1988) 121. 

\bibitem{CB89} C. Becchi and O. Piguet, {\sl Nucl. Phys.} {\bf B315}
(1989) 153. 

\bibitem{CB90} C. Becchi and O. Piguet, {\sl Nucl. Phys.} {\bf B347}
(1990) 596. 

\bibitem{BDV}  G. Bonneau, {\sl Nucl. Phys.} {\bf B221} (1983) 178 ;  G.
Valent, {\sl Nucl. Phys.} {\bf B238} (1984) 142 ;  G. Valent, {\sl Phys.
Rew.}  {\bf D30} (1984) 774 ;  F. Delduc and G. Valent, {\sl Nucl. Phys.}
{\bf B253} (1985) 494 ;  G. Bonneau, F. Delduc and G. Valent, {\sl Phys.
Lett.} {\bf 196B} (1987) 456.


\bibitem{FD99} F. Delduc, {\sl `` About holomorphic factorization"},
these proceedings.

\bibitem{GML60} M. GellMann and M. L\'evy, {\sl Nuovo Cimento} {\bf  16}
(1960) 705 .  

\bibitem{S57}  J. Schwinger, {\sl Ann. Phys. (N. Y.)} {\bf 2} (1957) 407.

\bibitem{NLS} S. Weinberg, {\sl Phys. Rev.}  {\bf 166} (1968) 1568 ; 
 J. Schwinger, {\sl Phys. Rev.}  {\bf 167} (1968) 1432 ; S. Coleman, J.
Wess and B. Zumino, {\sl Phys. Rev.}  {\bf 177} (1969) 2239 ; C. Callan,
S. Coleman, J. Wess and B. Zumino, {\sl Phys. Rev.}  {\bf 177} (1969)
2247. 

\bibitem{F} D. Friedan, {\sl Ann. Phys. (N. Y.)} {\bf 163} (1985) 318.

\bibitem{S1}   J. Schwinger, {\sl Phys. Rev.}  {\bf 167} (1968) 1432.

\bibitem{C1} J. M. Charap, {\sl Phys. Rev.}  {\bf D2} (1970) 1554.

\bibitem{Ring87} {\sl Renormalization of quantum fiel theories with
non-linear fiel transformations}, Lecture Notes in Physics $n^0 303\,,$
eds. P. Breitenlohner et {\sl al.} (Springer-Verlag, Berlin, 1988).  


\bibitem{BLASI99} A. Blasi, N. Maggiore, S. Sorella and Luiz C. Q. Vilar,
{\sl Phys. Rev } {\bf D59} (1999) 121701.
 
\bibitem{CB87} C. Becchi, {\sl ``The non-linear sigma model" }, in 
\cite{Ring87}. 

\bibitem{Meetz} K. Meetz, {\sl J. Math. Phys.} {\bf 10} (1969) 65 ; J.
Honerkamp,  {\sl Nucl. Phys.} {\bf B36} (1972) 130 ; G. Ecker and J.
Honerkamp,  {\sl Nucl. Phys.} {\bf B35} (1971) 481.

\bibitem{C2} J. M. Charap, \cite{C1} ; {\sl Phys. Rev.} {\bf D3} (1971)
1998 ;  J. Honerkamp and K. Meetz, {\sl Phys. Rev.} {\bf D3} (1970) 1996
;  I. S. Gerstein, R. Jackiw, B. W. Lee and S. Weinberg, {\sl Phys. Rev.}
{\bf D3} (1971) 2486. 

\bibitem{Valent1} G. Valent,  {\sl Phys. Rev.}  {\bf D30} (1984) 774.

\bibitem{BDV2} G. Bonneau, F. Delduc and G. Valent, {\sl Phys. Lett.} 
{\bf 196B} (1987) 456. 

\bibitem{BC87} A. Blasi and R. Collina, {\sl Nucl. Phys.} {\bf B285}
(1987) 204. 

\bibitem{PS82} O. Piguet and K. Sibold, {\sl Nucl. Phys.} {\bf B197}
(1982) 257 ; ibid. {\bf B248} (1984) 301. 


\bibitem{BM87} P. Breitenlohner, {\sl ``N=2 Supersymmetric Yang-Mills
theories in the Wess-Zumino gauge"}, in \cite{Ring87}. 

\bibitem{Kallosh} R. E. Kallosh, {\sl Nucl. Phys.} {\bf B141} (1978) 141.

\bibitem{BV} I. A. Batalin and G. A. Vilkovisky, {\sl Nucl. Phys.} {\bf
B234} (1984) 106. 

\bibitem{PS85} O. Piguet and K. Sibold, {\sl Nucl. Phys.} {\bf B253}
(1985) 269. 

\bibitem{Nicola} N. Maggiore,  {\sl Int. J. Mod. Phys.} {\bf A10} (1995)
3937 ; ibid. 3781.

\bibitem{B94} G. Bonneau, {\sl Helv. Phys. Acta} {67} (1994) 930 ; ibid.
954.

\bibitem{BM99} A. Blasi and N. Maggiore, {\sl J.H.E.P.} {\bf 9905} (1999)
8.

\bibitem{B1} G. Bonneau, {\sl Nucl. Phys.} {\bf B221} (1983) 178 ;  G.
Bonneau and F. Delduc, {\sl Nucl. Phys.} {\bf B266} (1986) 536.

\bibitem{BLGZJ} E. Brezin, J.C. Le Guillou and J. Zinn-Justin,  {\sl
Phys. Rev.} {\bf D14} (1976) 615. 

\bibitem{Maillet} H. de Vega and J.M. Maillet, {\sl Phys. Letters} {\bf
101B} (1981) 302.

\bibitem{BD2}  G. Bonneau, {\sl Phys. Letters} {\bf 133B} (1983) 341 ; G.
Bonneau and F. Delduc, {\sl Nucl. Phys.} {\bf B266} (1986) 536.


\bibitem{PLR} K. Pohlmeyer, {\sl Comm. Math. Phys.} {\bf 46} (1976) 207 ;
F. Lundt and T. Regge, {\sl Phys. Rev.} {\bf D14} (1976) 1524. 

\bibitem{BB90} G. Bonneau, {\sl Int. J. Mod. Phys.} {\bf A5} (1990) 3831.

\bibitem{GW96} J. Gomis and S. Weinberg, {\sl Nucl. Phys.} {\bf B469}
(1996) 473.


\bibitem{HKY} M. Harada, T. Kugo and K. Yamawaki, {\sl Prog. Theor.
Phys.} {\bf 91} (1994) 801.

\end{thebibliography}

\end{document}